\documentclass[%
 reprint,
superscriptaddress,
 amsmath,amssymb,
 aps,
floatfix,
]{revtex4-2}

\usepackage{latexsym}
\usepackage{graphicx}
\usepackage{dcolumn}
\usepackage{bm}
\usepackage{xcolor}

\begin{document}
\title{Sub-5-fs compression and synchronization of relativistic electron bunches \\enabled by a high-gradient $\alpha$-magnet and low-jitter photoinjector}

\author{Yining Yang}
\email{Email: yang-yn21@mails.tsinghua.edu.cn}

\affiliation{\mbox{ Department of Engineering Physics, Tsinghua University, Beijing, China 100084}}
\affiliation{\mbox{ Key Laboratory of Particle and Radiation Imaging (Tsinghua University), Ministry of Education, Beijing, China 100084}}

\author{Zhiyuan Wang}
\affiliation{\mbox{ Department of Engineering Physics, Tsinghua University, Beijing, China 100084}}
\affiliation{\mbox{ Key Laboratory of Particle and Radiation Imaging (Tsinghua University), Ministry of Education, Beijing, China 100084}}

\author{Peng Lv}
\affiliation{\mbox{ Department of Engineering Physics, Tsinghua University, Beijing, China 100084}}
\affiliation{\mbox{ Key Laboratory of Particle and Radiation Imaging (Tsinghua University), Ministry of Education, Beijing, China 100084}}

\author{Baiting Song}
\affiliation{\mbox{ Department of Engineering Physics, Tsinghua University, Beijing, China 100084}}
\affiliation{\mbox{ Key Laboratory of Particle and Radiation Imaging (Tsinghua University), Ministry of Education, Beijing, China 100084}}

\author{Pengwei Huang}
\affiliation{\mbox{ Department of Engineering Physics, Tsinghua University, Beijing, China 100084}}
\affiliation{\mbox{ Key Laboratory of Particle and Radiation Imaging (Tsinghua University), Ministry of Education, Beijing, China 100084}}

\author{Yanqing Jia}
\affiliation{\mbox{ Department of Engineering Physics, Tsinghua University, Beijing, China 100084}}
\affiliation{\mbox{ Key Laboratory of Particle and Radiation Imaging (Tsinghua University), Ministry of Education, Beijing, China 100084}}

\author{Zhuoxuan~Liu}
\affiliation{\mbox{ Department of Engineering Physics, Tsinghua University, Beijing, China 100084}}
\affiliation{\mbox{ Key Laboratory of Particle and Radiation Imaging (Tsinghua University), Ministry of Education, Beijing, China 100084}}

\author{Lianmin Zheng}

\affiliation{\mbox{ Department of Engineering Physics, Tsinghua University, Beijing, China 100084}}
\affiliation{\mbox{ Key Laboratory of Particle and Radiation Imaging (Tsinghua University), Ministry of Education, Beijing, China 100084}}
 
\author{Wenhui Huang}
\affiliation{\mbox{ Department of Engineering Physics, Tsinghua University, Beijing, China 100084}}
\affiliation{\mbox{ Key Laboratory of Particle and Radiation Imaging (Tsinghua University), Ministry of Education, Beijing, China 100084}}

\author{Pietro Musumeci}
\affiliation{\mbox{Department of Physics and Astronomy, University of California, Los Angeles, California, USA 90095}}

\author{Chuanxiang Tang}
\affiliation{\mbox{ Department of Engineering Physics, Tsinghua University, Beijing, China 100084}}
\affiliation{\mbox{ Key Laboratory of Particle and Radiation Imaging (Tsinghua University), Ministry of Education, Beijing, China 100084}}

\author{Renkai Li}
\email{Email: lirk@tsinghua.edu.cn} 

\affiliation{\mbox{ Department of Engineering Physics, Tsinghua University, Beijing, China 100084}}
\affiliation{\mbox{ Key Laboratory of Particle and Radiation Imaging (Tsinghua University), Ministry of Education, Beijing, China 100084}}

\date{Submitted 28 April, 2025; revised 31 July, 2025}

\begin{abstract}

Generating high-brightness relativistic electron bunches with few-femtosecond duration, while simultaneously achieving few-fs synchronization with ultrafast lasers, remains an outstanding challenge at the frontier of accelerator physics and ultrafast science. In this Letter, we present the beam physics and experimental demonstration of a new method that, for the first time, enables simultaneous control of bunch duration and synchronization with few-fs precision. Timing stabilization is achieved using a tailored high-gradient $\alpha$-magnet that optimizes the correlation between time of flight and momentum, together with a novel photocathode RF gun designed to suppress the effect of RF-to-laser timing jitter. Compression is realized by manipulating the time–momentum correlation in phase space, primarily through space-charge effects. Sub-5-fs rms bunch duration and synchronization are demonstrated. This method establishes a new regime in electron bunch control, unlocking new capabilities for ultrafast beam physics and applications.

\end{abstract}

\maketitle

Advancing the frontier of temporal control over relativistic electron bunches—including both duration and synchronization with ultrafast lasers—has long been a central thrust of beam physics and technology research, and a driver of transformative tools for ultrafast science. Such progress enables breakthroughs in ultrafast electron scattering~\cite{Mourou1982, Wang2003, Miller2014, Filippetto2022, King2005, Zewail2010, Armin2017}, allowing atomic motions to be tracked with unprecedented temporal resolution. It is also essential for advanced acceleration~\cite{Peralta2013, England2014, Sapra2020, Black2019, Chlouba2023, Nanni2015, Zhang2018, Hibberd2020, Xu2021, Tang2021, Esarey2009, Assmann2020, Wu2021, Lindstrom2024}, facilitating the injection of high-brightness electron bunches into laser-driven fields with enhanced precision and stability. 

Compression of electron bunch duration is achieved by first imparting a momentum chirp, followed by transport through a designated beamline with a well-defined correlation between particle time-of-flight (TOF) and momentum. In the RF velocity bunching approach~\cite{Anderson2005, Oudheusden2010, Li2011, Gao2012, Chatelain2012, zeitler2015, Maxson2017, CornellUED2022, Denham2023}, an RF field imposes a positive momentum chirp, causing trailing particles to gain higher momentum and catch up with the bunch head during the drift, thereby compressing the bunch duration. Although sub-10 fs durations have been demonstrated, synchronization with ultrafast lasers remains limited to several tens of fs due to fluctuations in RF amplitude and RF-to-laser timing. 

THz compression schemes employ THz fields generated by ultrafast lasers to impart a momentum chirp to electron bunches, compressing their duration and stabilizing TOF~\cite{Kealhofer2016, Ehberger2019, Snively2020, Zhao2020}. Electron bunch durations and synchronization with lasers at the few-tens-of-fs level have been demonstrated. Further development is needed to overcome limitations from small structure apertures, transverse–longitudinal field coupling, and THz waveform nonlinearity to enhance reliability and performance. The demand for substantial laser energy to generate strong, stable THz fields also poses challenges for kHz or higher repetition rates. Several laser-based concepts for few- and sub-fs isolated bunch control have been proposed and await experimental validation~\cite{ChengLi2022, Guo23, Cheng2025}. 

\begin{figure*}[!t]
    \centering
    \includegraphics[width=0.96\linewidth]{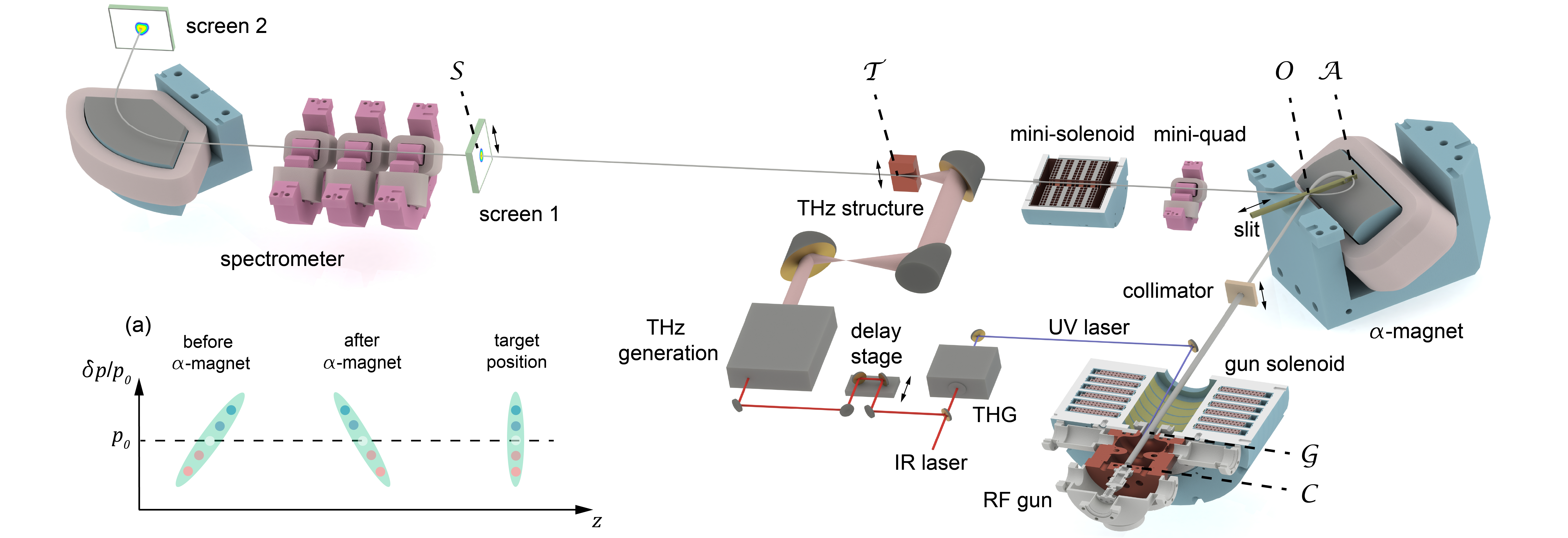}
    \caption{Schematic of the beamline layout. (a) Evolution of the electron bunch's longitudinal phase space along the beamline.}
    \label{fig:f1}
\end{figure*}

An alternative approach to enhanced temporal control employs a static magnetic transport line with a defined TOF–momentum correlation~\cite{Craievich2013, He2015, Faure2016, Pompili2016, Inoue2020}, expressed as \mbox{$R_{56}=\delta z/(\delta p/p)$} in beam transport matrix formalism. For example, a double-bend achromat (DBA) can suppress momentum-induced TOF jitter~\cite{Kim2019, Qi2020} and compress bunches with initial chirps from space-charge effects and RF fields in a photocathode RF gun, eliminating the need for additional RF or THz fields. However, the momentum–TOF jitter coupling in conventional photocathode RF guns, combined with DBA's properties and complexity, has made the designed performance challenging to realize in practice and yet to be demonstrated~\cite{Kim2020, Xu2024}.

Despite extensive efforts using various methods, the best combined temporal control precision to date—reported by DBA schemes—remains well above the 10-fs barrier, a critical and long-standing challenge posed by next-generation ultrafast electron scattering, advanced acceleration, and the continued pursuit of limits in beam physics and control. Even marginal improvements with existing methods have stalled for over half a decade, and viable paths toward single- and sub-fs precision remain entirely elusive.

In this Letter, we introduce the concept, investigate the underlying beam physics, and report the first experimental demonstration of a new method that simultaneously achieves few-fs compression and TOF stabilization of high-brightness MeV electron bunches. This enhanced temporal control is realized by unifying two innovative, specially tailored elements: a high-gradient $\alpha$-magnet~\cite{Enge1963, Borland1991}, which tunes the TOF-momentum correlation with negligible transverse dispersion, and a novel-geometry S-band photocathode RF gun~\cite{Li2014, Pirez14, Song2022, Xu2025} with a shortened first cell that suppresses TOF jitter arising from RF-to-laser timing fluctuations. We demonstrate that both the bunch duration and TOF jitter relative to ultrafast lasers are simultaneously reduced well below the long-standing 10-fs barrier, reaching sub-5-fs rms, as characterized by laser-driven THz streaking. The transverse emittance of the compressed bunches was measured, confirming effective brightness preservation. The beam control, phase-space evolution, and characterization results are discussed.

$\alpha$-magnets have been used for decades primarily with thermionic RF guns for bunch compression, without precise synchronization with ultrafast lasers~\cite{Tang1999, pal2004, Lee2007, Gu2008, miyahara:ipac10, yang2011, Shin2018, lee:napac2019, Rajabi2019, marzouk:linac2024, pires:ipac23}. In contrast, employing an $\alpha$-magnet to control the TOF relative to ultrafast lasers with a photocathode RF gun is a viable yet previously unexplored concept. This novel intended use necessitates pushing the $\alpha$-magnet design to unprecedented strength. To meet this requirement, we developed one with a $\sim$15 T/m gradient, far exceeding the typical 5 T/m or lower level, representing the highest-gradient electromagnetic $\alpha$-magnet reported to date.


While DBA and triple-bend achromat (TBA)~\cite{Hounsell2025} are broadly applicable across a wide energy range, at the few-MeV level the $\alpha$-magnet offers several key advantages. Its simplicity—a single element compared to multiple magnets in DBA/TBA—substantially facilitates beam tuning and optimization. It provides a tunable $R_{56}$ for flexible TOF control, compared to a fixed $R_{56}$ in DBA or the added complexity in TBA. Its high-order optics are favorable, with $T_{566} = -R_{56}/4$ at the low few-cm level. The short path length (0.1–0.2 m), compared to the meter-scale transport in DBA/TBA, also mitigates space-charge effects. Combined with a novel photocathode RF gun, the $\alpha$-magnet enables a simple, tunable, compact, and effective approach to few-fs beam control.

The layout of the scheme is depicted in Fig.~\ref{fig:f1} and implemented at FORTRESS (Facility Of Relativistic Time-Resolved Electron Sources and Scattering) at Tsinghua University~\cite{Peng2024}. The electron source is a novel 1.4-cell S-band photocathode RF gun. A Ti:Sapphire ultrafast laser delivers 9 mJ, 800 nm pulses split 1:9 between UV generation for the photocathode and THz generation for electron bunch characterization. Electron bunches exit the gun with $\sim$3 MeV kinetic energy, following an $\alpha$-shaped trajectory in the $\alpha$-magnet, and propagate further downstream. The intersection point of the $\alpha$ trajectory $\mathcal{O}$ is 0.95 m from the photocathode. A THz streaking structure is placed 1.34 m from $\mathcal{O}$, and a transverse beam profile screen is located 3.22 m downstream. A spectrometer, consisting of a quadrupole triplet and a dipole, is used in combination with THz streaking to characterize the longitudinal phase space (LPS).

\textbf{TOF control}. We aim to stabilize the TOF of electron bunches from the photocathode $\mathcal{C}$ to the target position $\mathcal{T}$, despite fluctuations primarily induced by RF amplitude and RF-to-laser timing jitters. A novel photocathode RF gun plays a pivotal role in this method. Unlike conventional designs, the new 1.4-cell geometry tailors the longitudinal beam dynamics so that the RF phases for maximum momentum gain and minimal TOF closely overlap. As a result, operating near this optimal phase substantially suppresses the impact of RF-to-laser timing jitter on both the bunch momentum and TOF from $\mathcal{C}$ to the gun exit $\mathcal{G}$. The TOF within the gun then depends solely on the RF amplitude, which directly correlates with the bunch momentum, enabling the definition of a characteristic gun $R_{56}^{\mathcal{CG}}$. The novel gun is critical for disentangling the coupling between momentum and TOF jitters, allowing significantly cleaner and more effective TOF control, and thereby achieving enhanced precision.

The $R_{56}$ for the drift sections $\mathcal{GO}$ and $\mathcal{OT}$ are $L/\gamma^2$, where $L$ is the path length. Both the RF gun and drifts contribute positive $R_{56}$ values at the cm level. In first-order optics, electrons follow $\alpha$-shaped trajectories in an $\alpha$-magnet, as illustrated in Fig.~\ref{fig:f1}, with identical, momentum-independent entrance and exit paths that intersect at point $\mathcal{O}$. Higher-momentum electrons travel along longer trajectories than lower-momentum ones, resulting in {$R_{56}^\mathcal{\alpha}(\text{cm})=-9.6\sqrt{\gamma\beta/g(\text{T/m})}$}, where $g$ is the magnetic field gradient. Under our beam conditions, higher-order optics of the $\alpha$-magnet and space-charge effects have negligible influence on TOF. The negative sign and tunability of $R_{56}^\mathcal{\alpha}$ are crucial for counteracting the positive $R_{56}$ contributions from the gun and drift sections, thereby minimizing TOF fluctuations.

Stabilization of TOF is achieved by tuning the momentum-induced timing difference, which depends on {$R^{tot}_{56}=R_{56}^{\mathcal{CG}}+R_{56}^{\mathcal{GO}}+R_{56}^{\mathcal{\alpha}}+R_{56}^{\mathcal{OT}}$}, to approach zero. We designed and implemented the beamline layout based on this criterion, and adjusted the $\alpha$-magnet during experiments to precisely control $R^{tot}_{56}$.

\textbf{Compression control}. Full compression of the bunch duration is achieved when electrons in an initially chirped and elongated bunch are mapped into a minimal temporal spread through beam transport, expressed as \mbox{$1+hR_{56}=0$}, where \mbox{$h = (\delta p / p) / \delta z$} represents the initial momentum chirp. This criterion is satisfied when the bunch momentum remains constant and space-charge effects are negligible. We adopt \mbox{$1 + h^{\mathcal{O}}(R_{56}^\alpha + R_{56}^\mathcal{OT}) = 0$} as the guideline for optimizing the compression process. The required momentum chirp at the entrance of the $\alpha$-magnet, $h^\mathcal{O}$, is primarily developed by space-charge effects, which depend on the evolving bunch charge density shaped by its initial distribution at photoemission, as well as subsequent acceleration and transverse focusing. Space-charge effects are particularly pronounced at low bunch energies and high densities near the photocathode, and diminish substantially as the bunch is accelerated to relativistic energies. With settings optimized for minimal TOF jitter, the drive laser is adjusted to tailor the initial bunch spot size, duration, and charge, tuning $h^{\mathcal{O}}$ over a range sufficient for full compression.



\begin{table}[!hbt]
    \caption{\label{tab:beamparameters} Main electron bunch and apparatus parameters}
    \begin{ruledtabular}
        \begin{tabular}{lcdr}
            \textrm{Parameters} & \multicolumn{1}{c}{\textrm{Value}} & \multicolumn{1}{c}{\textrm{Unit}} \\
            \colrule
            RF gun gradient              & 75.8        & \multicolumn{1}{c}{MV/m} \\
            Bunch kinetic energy            & 2.92         & \multicolumn{1}{c}{MeV} \\
            Initial bunch length (FWHM)    & 170          & \multicolumn{1}{c}{{fs}} \\
            Initial bunch spot size (rms)          & 53          & \multicolumn{1}{c}{$\mu$m} \\
            Initial bunch charge $Q_\text{ini}$            & 0.3$\sim$3           & \multicolumn{1}{c}{pC} \\
            
        \end{tabular}
    \end{ruledtabular}
\end{table}

\textbf{Experimental results.} Electron bunches generated by the RF gun at 50 Hz are focused by the gun solenoid, collimated by a 200 $\mu$m diameter aperture located at 0.6 m, and injected into the $\alpha$-magnet. The $\alpha$-magnet's gradient is tunable from a lower bound of 6.85 T/m, where the apex of the $\alpha$-shaped trajectory $\mathcal{A}$ reaches the vacuum chamber wall, up to 14.75 T/m. It is configured to produce an angle of $\sim$81.4$^\circ$ between the entrance and exit paths. A downstream mini quadrupole magnet and solenoid are used to compensate for the $\alpha$-magnet's asymmetric transverse focusing and to restore and control a round transverse profile. THz streaking at $\mathcal{T}$ is employed for temporal characterization of the electron bunches~\cite{Fabianska2014, Zhao2018, Li2019, Baek2021} (see Appendix~\ref{sec:appendix:thzstreaking}). The main beam and machine parameters are summarized in Table~\ref{tab:beamparameters}. 

\begin{figure}[h]
    \centering
    \includegraphics[width=1\linewidth]{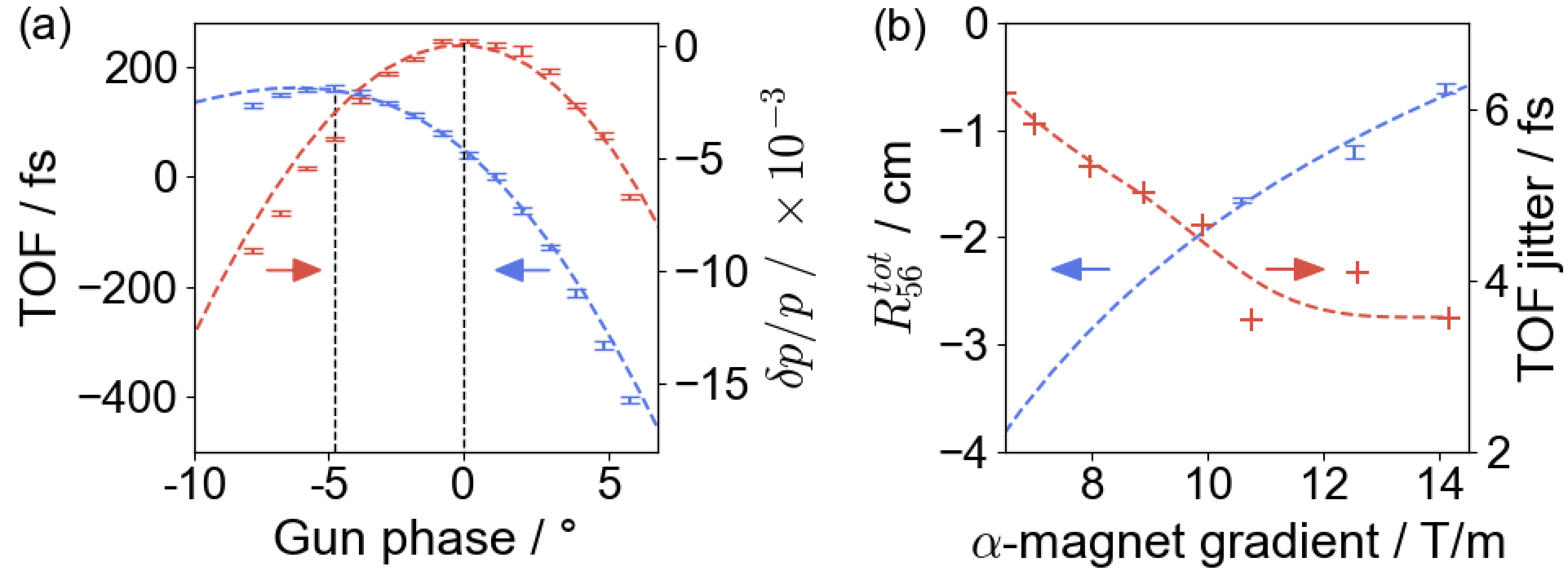}
    \caption{(a) Measured bunch TOF (blue) and relative momentum deviation $\delta p/p$ (red) with error bars, as a function of gun launch phases. Simulation results are shown as dashed lines. (b) Measured $R_{56}^{tot}$ (blue, with error bars) and TOF jitter (red) for various $\alpha$-magnet gradients.}
    \label{fig:ekandtof}
\end{figure}

We first characterized the bunch momentum and TOF at different gun launch phases $\phi$, as shown in Fig.~\ref{fig:ekandtof}(a). The launch phases for maximum momentum gain, defined as $0^\circ$ here, and for maximum TOF, differ by only $4.8^\circ$, consistent with the gun design and beam dynamics simulations performed using the General Particle Tracer (GPT) code~\cite{gpt}. For subsequent beam optimization and measurements, we operated at 0$^\circ$, where the $\phi$-TOF correlation is 39.3 fs/degree, or 0.039 when converting RF phase to time, demonstrating effective suppression of TOF fluctuations induced by RF-to-laser timing jitter. With a routine RF-to-laser timing jitter of 26 fs rms, retrieved from the bunch momentum and TOF stability measurements prior to installation of the $\alpha$-magnet~\cite{Peng2024}, it contributes only approximately 1.0 fs rms to overall TOF fluctuation when combined in quadrature with other sources. This marks the first experimental demonstration of a gun geometry that effectively suppresses RF-to-laser timing-jitter–induced TOF fluctuations. 

We characterized $R_{56}^\text{tot}$ at selected $\alpha$-magnet gradients by varying the RF amplitude near the nominal setting and correlating the measured TOF with momentum change. This measurement was enabled by one of the key advantages of an $\alpha$-magnet—its tunability. The results agree well with simulations, as shown in Fig.~\ref{fig:ekandtof}(b). TOF jitter, measured at various $\alpha$-magnet gradients and $R_{56}^\text{tot}$ values, levels off at $\sim$4 fs rms for gradients $>10$ T/m or $|R_{56}^\text{tot}| < 2$ cm, as shown in Fig.~\ref{fig:ekandtof}(b). 

Residual TOF jitter may arise from vibrations of optical elements and pointing jitter in the laser transport lines for the cathode drive and THz branches, including THz waveform fluctuations~\cite{Kuttruff21}. An additional source is variation in the $\alpha$-magnet field strength, which affects the path length according to $dS_{\alpha}(\text{cm})=-9.6\sqrt{\beta\gamma/g(\text{cm})}\frac{\delta B}{B}$, where a 10 ppm fluctuation in $\frac{\delta B}{B}$ leads to a 2.6 fs rms TOF jitter. An $\alpha$-magnet gradient of 10.7 T/m, corresponding to the measured {$R_{56}^\text{tot} = -1.67$ cm}, was employed for the measurements in Fig.~\ref{fig:ekandtof}(a) and for subsequent bunch compression. This setting effectively minimizes TOF jitter—the measured bunch momentum stability, primarily influenced by RF amplitude fluctuations, is $2.6 \times 10^{-5}$, corresponding to a TOF jitter of 1.4 fs rms—and also relaxes the required initial bunch momentum chirp for optimal compression.

\begin{figure}[h]
    \centering
    \includegraphics[width=1\linewidth]{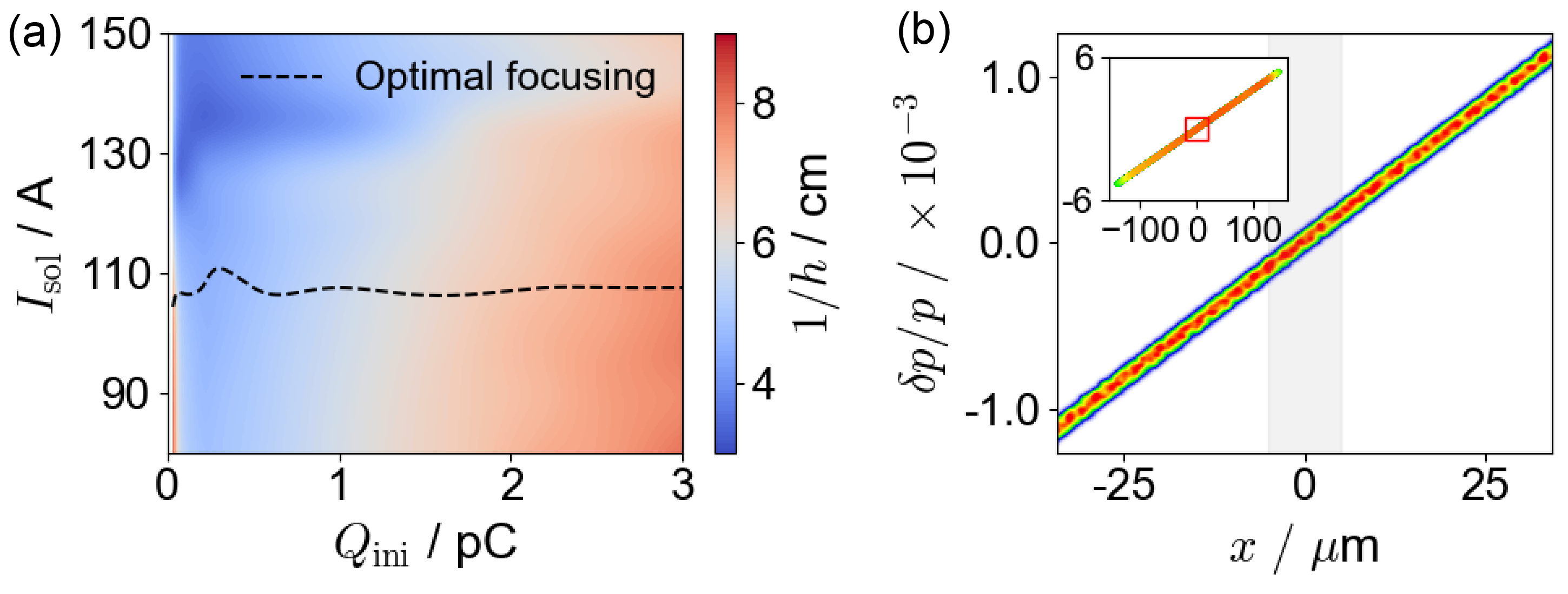}
    \caption{(a) Simulated bunch chirp at the entrance of the $\alpha$-magnet for different initial charges $Q_\text{ini}$ and gun solenoid current $I_\text{sol}$. The optimal solenoid current to focus the beam onto the slit is shown by the dashed line. (b) Simulated \mbox{$x$-$\delta p/p$} correlation on the slit plane. Shaded area indicates the slit opening. Inset shows the overall $x$-$\delta p/p$ correlation.}
    \label{fig:chirpandslit}
\end{figure}

Compression of the electron bunch is accomplished by generating a controllable chirp that matches the beamline's transport properties. The cathode drive laser featured a quasi-uniform transverse profile with a size of 53 $\mu$m rms, and its pulse duration was measured to be \mbox{170 fs FWHM}. The initial bunch charge $Q_\text{ini}$, which is continuously tunable, was adjusted to control the bunch chirp. $Q_\text{ini}$ was relatively high, up to the pC level, facilitating the generation of a sufficient chirp. The collimator transmitted the transverse core of the bunch at the highest brightness and reduced the transverse emittance to the few-nm-rad level, a typical value required for high $q$-resolution UED experiments. The dependence of $h^\mathcal{O}$ on $Q_\text{ini}$ and the transverse focusing, modeled using experimental parameters, is shown in Fig.~\ref{fig:chirpandslit}(a). Simulation results confirm that when $h^\mathcal{O}$ matches $R_{56}^\alpha+R_{56}^\mathcal{OT}$, the bunch undergoes near-optimal compression, verifying that after collimation space-charge effects are substantially reduced and exert only a minor influence on bunch chirp evolution during transport and compression.  


\begin{figure*}[!t]
    \label{fig:LPS} 
    \centering
    \begin{minipage}{0.70\textwidth}
        \centering
        \includegraphics[width=1\linewidth]{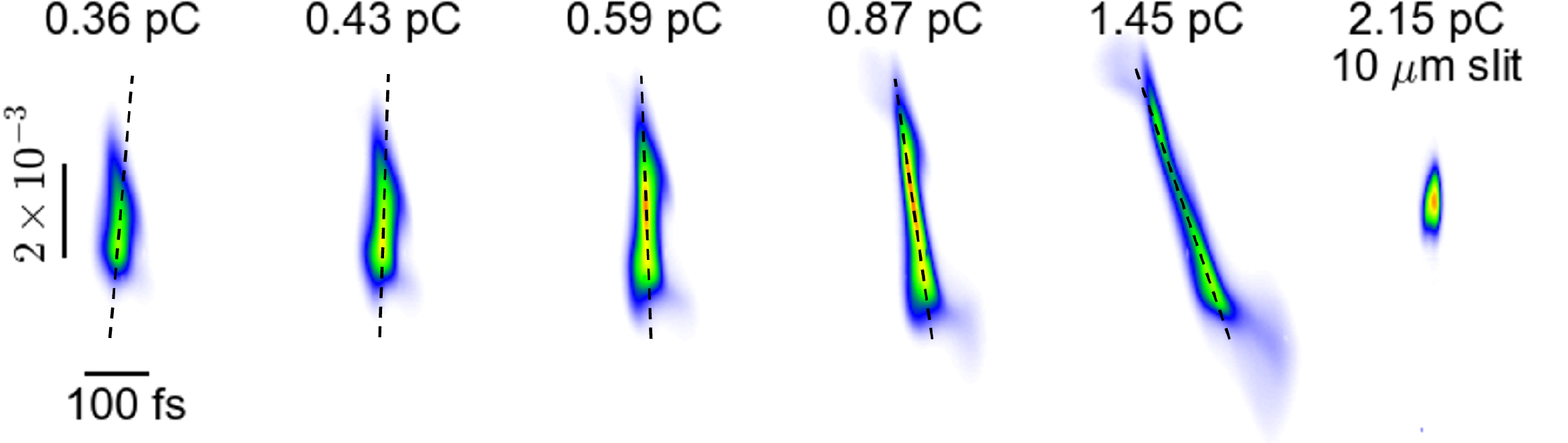}
    \end{minipage}\hfill
    \begin{minipage}{0.28\textwidth}
        \caption{ \label{fig:LPS} Measured bunch LPS distributions for different initial charges. Dashed lines indicate the momentum chirp. For the right-most panel a 10 $\mu$m slit was used for LPS selection, while no slit for the others.}
    \end{minipage}
\end{figure*}

The linearity of the $t$-$\delta p/p$ correlation is crucial for shaping the final LPS distribution at the longitudinal focus. Nonlinearities—arising from space-charge effects and higher-order beam transport—limit the minimum bunch duration and are most pronounced at the bunch head and tail, as confirmed by GPT simulations. The three built-in space-charge models in GPT—the 3D mesh, tree-code, and point-to-point algorithms—produce consistent results after a systematic convergence check~\cite{3Dmesh2004}. Compression is enhanced by selecting the central portion of the LPS. A movable collimation slit was installed along the $\mathcal{OA}$ direction in the vertical mid-plane of the $\alpha$-magnet, where the transverse position and particle momentum are tightly correlated. Fig.~\ref{fig:chirpandslit}(b) shows the simulated $x$-$\delta p/p$ distribution at the slit location, with a correlation coefficient of 3.2 cm, in good agreement with the $\alpha$-magnet's intrinsic value \mbox{$R_{16}\text{(cm)} \approx 3.75 \sqrt{\gamma\beta/g\text{(T/m)}} = 3.0$ cm}. The gun solenoid is tuned to focus the transverse bunch size at the slit, minimizing the $x$-spread for each $p$ and maximizing the $t$-$\delta p/p$ correlation of the transmitted bunch. This architecture—combining transverse focusing, the $\alpha$-magnet’s optical properties, and the slit—provides a clear and effective approach to precise phase-space selection.

By utilizing a 10 $\mu$m slit for LPS selection and tuning $Q_\text{ini}$ to optimize the chirp for compression, a bunch duration of $3.47\pm 0.33$ fs rms was achieved with a TOF jitter of 4.34 fs rms, yielding a combined control precision of $5.56\pm0.21$ fs rms—a breakthrough with a $\sim$3-fold enhancement over the previous best, in a regime where even incremental gains are challenging. The normalized transverse emittance of the compressed bunch was 7.2 nm·rad, measured using the single-slit scan method. The effective preservation of emittance during compression indicates negligible growth from $\alpha$-magnet's higher-order optics and space-charge effects, as confirmed by GPT simulations. The bunch entered the $\alpha$-magnet with a charge of 45 fC, and 1.6 fC—or $1.0\times10^4$ electrons—was transmitted through the 10 $\mu$m slit, producing a notably high 5D brightness of $B_\textrm{5D}=2I/\epsilon_n^2=7.1\times10^{15}~\textrm{A}/(\textrm{m}~\textrm{rad})^2$. The momentum spread was measured to be $(5.7\pm 0.37)\times10^{-4}$ rms, with a centroid jitter of $2.6\times 10^{-5}$ rms. Simulations showed that increasing the slit width leads to proportional growth in bunch charge and energy spread, with minimal change in the compressed bunch duration—requiring chirp refinement—until nonlinearities at the bunch head and tail limit further compression. The optimal $Q_\text{ini}$ for minimal bunch duration was 2.15 pC. Experimentally, varying $Q_\text{ini}$ between 2 and 3 pC resulted in only minor changes in bunch duration (3.47–3.63 fs) and bunch charge (1.66–1.46 fC), demonstrating high robustness against charge variations, enabling stable operation under realistic conditions. 


Snapshots of the LPS distributions, obtained by transporting the THz-streaked bunches through the spectrometer, provide valuable insights into the space-charge-driven compression dynamics, as shown in Fig.~\ref{fig:LPS}. As $Q_\text{ini}$ increases, leading to a growing momentum chirp, the rotation of the LPS can be precisely controlled and clearly visualized, revealing the transition from under- to over-compression. The optimal bunch duration without a slit was \mbox{$14.3 \pm 0.33$ fs rms}, achieved with a $Q_\text{ini}$ of 0.59 pC, corresponding to a chirp in good agreement with the $h^\mathcal{O}$ value required by the $\alpha$-magnet setting and drift distance of $R_{56}^\alpha+R_{56}^\mathcal{OT}=-4.61$ cm, as shown in Fig.~\ref{fig:chirpandslit}(a). Another notable feature in the LPS snapshots is the formation of halos at the bunch edges as the charge increases, while the core remains linear and tends to compress in the direction perpendicular to the chirp. These findings underscore the effectiveness of using a slit to select the bunch core, enabling flexible control over the bunch duration without compromising the TOF fluctuation.  

\begin{figure}[h]
    \centering
    \includegraphics[width=1\linewidth]{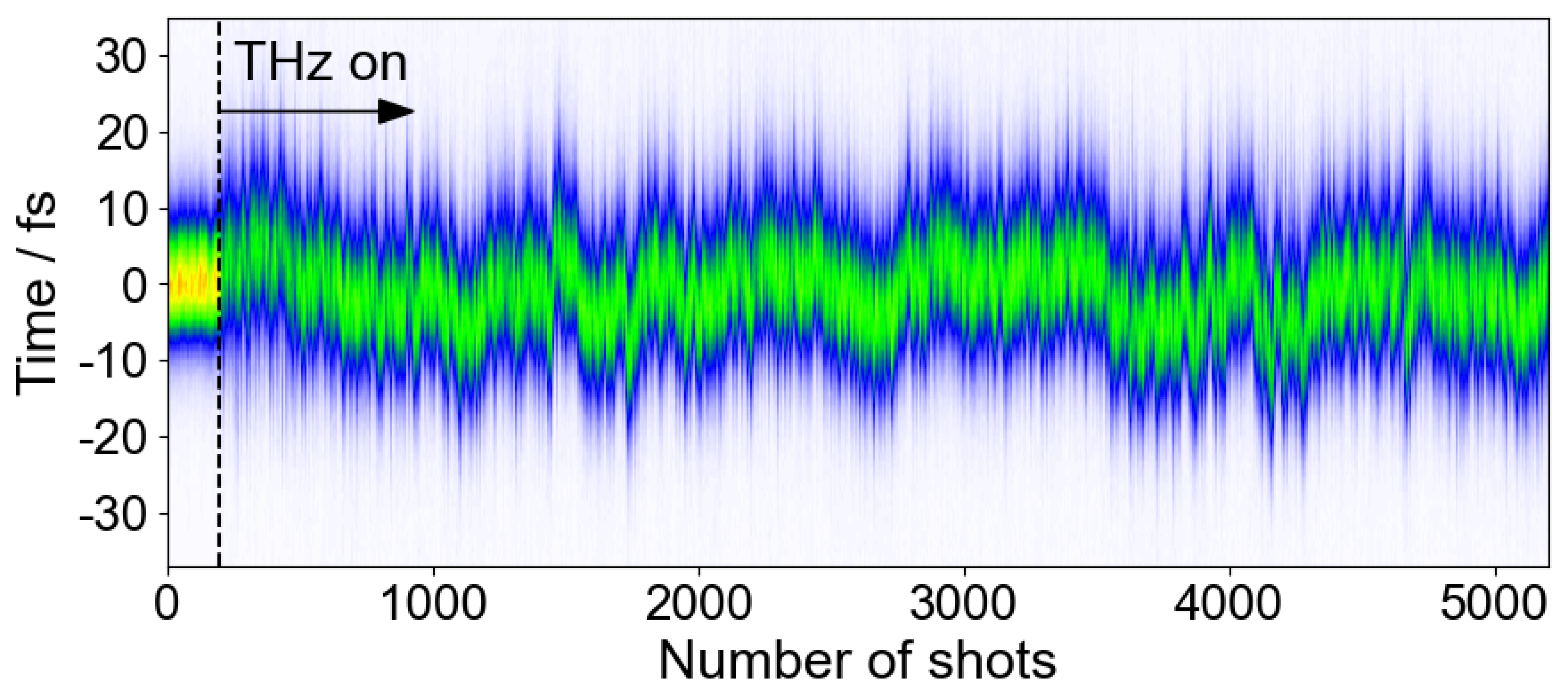}
    \caption{5000 consecutively recorded single-shot THz-streaked bunch profiles, with 200 shots recorded with THz off shown to the left of the dashed line for reference. Note that the unstreaked bunch profile must be subtracted in quadrature from the streaked profile to obtain the bunch duration values.}
    \label{fig:longterm}
\end{figure}

In addition to the flexibility and precision in beam control, the simplicity of this scheme’s apparatus ensures high robustness. It utilizes a single RF structure—the RF gun—requiring only one RF power source with high amplitude and phase-to-laser timing stability. TOF control is achieved primarily by a single electro-optical element—the $\alpha$-magnet, powered by a state-of-the-art, high-stability power supply. Fig.~\ref{fig:longterm} presents 5000 single-shot images of THz-streaked electron bunches, recorded consecutively over 1200 seconds (limited by camera readout rate), demonstrating long-term bunch stability. 

In conclusion, we present the physics and first experimental demonstration of a novel beam manipulation method that, with a minimal set of acceleration and electro-optical elements, offers an elegant, robust, and high-performance approach to significantly advancing the temporal control of bright relativistic electron bunches. This method enables deeper beam-dynamics insights and establishes a higher-precision phase-space control framework that pushes beyond previous approaches. We demonstrate simultaneous control of bunch duration and synchronization with ultrafast lasers to sub-5 fs, substantially surpassing the critical and long-standing 10-fs barrier and marking a paradigm shift in precise temporal control. This method readily extends to MeV, high-repetition-rate electron sources. Electron bunches controlled at this level will drive progress in advanced acceleration and ultrafast science, while extending the frontiers of beam physics and ultrafast technologies, e.g., paving the way toward attosecond-level control of bright electron bunches for future discoveries.

The authors thank Y. Du, W. Wan, T. Xu for insightful discussions, and R. Chen, X. Liu, X. Xu, L. Yun, N. Zha, Y. Zhu for technical support. This work was partially supported by National Key Research and Development Program of China under Grant No. 2022YFA1603400, 2024YFF1500900, and Tsinghua University Initiative Scientific Research Program under Grant No. 20197050028. P. M. was partially supported by the National Science Foundation under Grant No. DMR-1548924.

Y. Yang, Z. Wang and P. Lv contributed equally to this work. 

\appendix

\section{THz streaking}
\label{sec:appendix:thzstreaking}

\begin{figure}[h]
    \centering
    \includegraphics[width=1\linewidth]{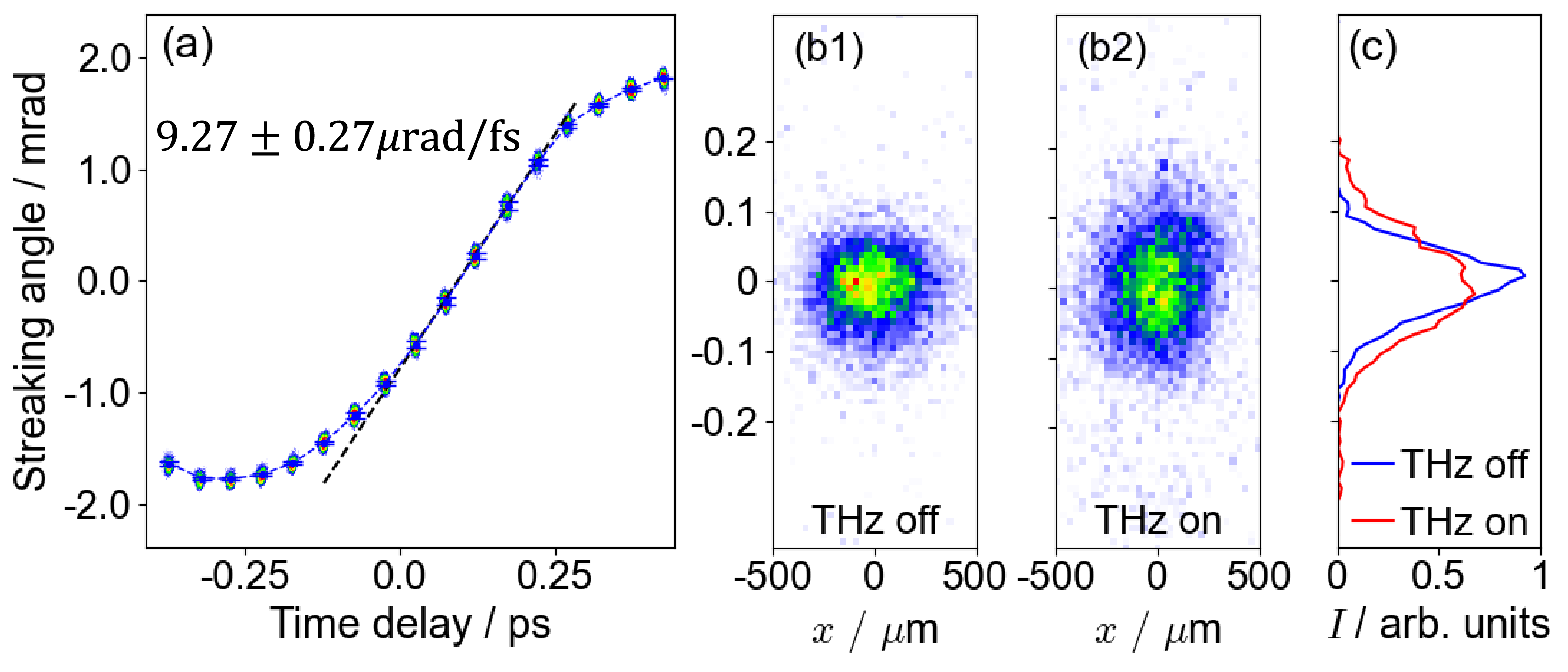}
    \caption{(a) Measured THz streaking calibration curve. Single-shot bunch profiles for (b1) THz off and (b2) THz on, and (c) corresponding intensity $I$ lineouts projected onto the streaking direction axis.}
    \label{fig:thzstreaking}
\end{figure}

THz streaking was used to characterize the temporal distribution of the electron bunches. A THz pulse is generated through optical rectification of the 800 nm laser pulse in a LiNbO$_3$ crystal, which is collinearly injected with the electron bunches into a taper-enhanced THz structure and imparts a time-dependent vertical deflection to the electrons. The streaked electron bunch profiles are recorded by screen 1, revealing the bunch duration and temporal jitter relative to ultrafast pulses with which the THz fields are inherently synchronized. The calibration curve of the THz streaking is shown in Fig.~\ref{fig:thzstreaking}(a). The quasi-linear region with a streaking strength of $K=9.27\pm0.27~\mu$rad/fs is subsequently used for highest-resolution temporal characterization. The timing accuracy $t^\text{res}_0$, defined as $\delta y'_0/K$ where $\delta y'_0$ is the vertical pointing jitter with THz off, is 0.51 fs rms. The minimal resolvable bunch duration $\sigma_t^\text{res}$ is determined as when the streaked bunch size is larger than the unstreaked bunch size $\sigma_y$ by 3 times the fluctuation of the unstreaked bunch size $\delta\sigma_y$, and is approximately $\sqrt{6\sigma_y\delta \sigma_y}/KL^\mathcal{TS}$. The measured $\sigma_y$ and $\delta \sigma_y$ are 172.4 and 5.6 $\mu$m, respectively, yielding a $\sigma_t^\text{res}=2.55$ fs rms. 

\bibliography{revision}
\end{document}